\documentclass[aip,numerical,amsmath,amssymb,reprint]{revtex4-1}

\draft 

\usepackage[caption=false]{subfig}
\usepackage{graphicx}

\usepackage{threeparttable}
\usepackage{multirow}

\usepackage{gensymb}
\usepackage{nicefrac}

\usepackage{algorithmicx}
\usepackage[linesnumberedhidden,ruled,vlined]{algorithm2e}

\begin{document}


\title{Fast dose optimization for rotating shield brachytherapy} 



\author{Myung Cho}
\affiliation{Department of Electrical and Computer Engineering, University of Iowa, 4016 Seamans Center, Iowa City, IA 52242}

\author{Xiaodong Wu}
\email[Author to whom correspondence should be addressed. Electronic mail: ]{xiaodong-wu@uiowa.edu}
\affiliation{Department of Electrical and Computer Engineering, University of Iowa, 4016 Seamans Center, Iowa City, IA 52242}
\affiliation{Department of Radiation Oncology, University of Iowa, 200 Hawkins Drive, Iowa City, Iowa, 52242}

\author{Hossein Dakhah}
\affiliation{Department of Biomedical Engineering, University of Iowa, 1402 Seamans Center, Iowa City, Iowa 52242}

\author{Jirong Yi}
\affiliation{Department of Electrical and Computer Engineering, University of Iowa, 4016 Seamans Center, Iowa City, IA 52242}

\author{Ryan T. Flynn}
\affiliation{Department of Radiation Oncology, University of Iowa, 200 Hawkins Drive, Iowa City, Iowa, 52242}

\author{Yusung Kim}
\affiliation{Department of Radiation Oncology, University of Iowa, 200 Hawkins Drive, Iowa City, Iowa, 52242}

\author{Weiyu Xu}
\affiliation{Department of Electrical and Computer Engineering, University of Iowa, 4016 Seamans Center, Iowa City, IA 52242}


\date{15 April 2017}

\begin{abstract}
\noindent\textbf{Purpose:} To provide a fast computational method, based on the proximal graph solver (POGS) - a convex optimization solver using the alternating direction method of multipliers (ADMM), for calculating an optimal treatment plan in rotating shield brachytherapy (RSBT). RSBT treatment planning has more degrees of freedom than conventional high-dose-rate brachytherapy due to the addition of emission direction, and this necessitates a fast optimization technique to enable clinical usage. \\
\textbf{Methods:} The multi-helix RSBT (H-RSBT) delivery technique\cite{dadkhah2015multihelix} was considered with five representative cervical cancer patients. Treatment plans were generated for all patients using the POGS method and the previously considered commercial solver IBM ILOG CPLEX\cite{cplex}. The rectum, bladder, sigmoid colon, high-risk clinical target volume (HR-CTV), and HR-CTV boundary were the structures considered in our optimization problem, which is called the asymmetric dose-volume optimization with smoothness control. Dose calculation resolution was $1 \times 1 \times 3$ $mm^3$ for all cases. The H-RSBT applicator has 6 helices, with $33.3\; mm$ of translation along the applicator per helical rotation and $1.7\; mm$ spacing between dwell positions, yielding 17.5\degree emission angle spacing per $5\; mm$ along the applicator. \\
\textbf{Results:} For each patient, HR-CTV $D_{90}$, HR-CTV $D_{100}$, rectum $D_{2cc}$, sigmoid $D_{2cc}$, and bladder $D_{2cc}$ matched within $1\%$ for CPLEX and POGS methods. Also, we obtained similar EQD2 figures between CPLEX and POGS methods. POGS was around 18 times faster than CPLEX. Over all patients, total optimization times were 32.1-65.4 seconds for CPLEX and 2.1-3.9 seconds for POGS.\\
\textbf{Conclusions:} POGS substantially reduced treatment plan optimization time around 18 times for RSBT with similar HR-CTV $D_{90}$, organ at risk (OAR) $D_{2cc}$ values, and EQD2 figure relative to CPLEX, which is significant progress toward clinical translation of RSBT. POGS is also applicable to conventional high-dose-rate brachytherapy.
\end{abstract}


\maketitle 

\section{INTRODUCTION}
\label{sec:intro}
High-dose-rate brachytherapy (HDR-BT) involves placing a radiation source inside of or adjacent to a target organ, i.e., tumor. Conventional HDR-BT uses an unshielded brachytherapy source with a radially-symmetric dose distribution\cite{potter2007clinical,kirisits2006vienna}, which limits the intensity modulation capacity of the approach. Rotating-shield brachytherapy (RSBT) has a rotating radiation-attenuating shield around a brachytherapy source. The RSBT concepts for single-catheter treatment\cite{ebert2002possibilities} and multi-catheter treatment\cite{ebert2006potential} were introduced by Ebert in 2002 and 2006 respectively.

In the multi-helix RSBT (H-RSBT) treatment, a radiation source travels inside a brachytherapy applicator having helical keyways. While moving along the applicator for a given keyway, the partial shield rotates around the radiation source simultaneously. In traveling along each keyway, the radiation source stops at designated locations called dwell positions. By adjusting the distance between adjacent dwell positions, the rotation angle of the partial shield is determined accordingly. Intensity modulated dose distributions can be delivered to the target with reduced dose exposure to non-target organs by controlling the treatment time in an optimal manner for each dwell position. Hence, it is reported that a radiation source with rotating shields can deliver more conformal dose distributions than an unshielded radiation source.\cite{dadkhah2015multihelix}

HDR-BT treatment plans are often generated using inverse planning tools\cite{lessard2001inverse,akimoto2006anatomy,baltas2009influence,citrin2005inverse,dewitt20053d,manikandan2013role,jacob2008anatomy}. Based on the given clinical prescription, various optimization problems were introduced previously ranging from minimizing treatment time under restrictions\cite{kneschaurek1999volume,renner1990algorithm} to minimizing dose error\cite{lessard2001inverse,manikandan2013role,jacob2008anatomy,alterovitz2006optimization}. Inverse planning by simulated annealing (IPSA)\cite{lessard2001inverse} is a well known method to optimize the dose volume histogram (DVH) directly with given constraints. The BrachyVision treatment planning system (Varian Medical System Inc., Palo Alto, CA) uses this type of DVH-based optimization algorithm.\cite{manikandan2013role}

Unlike the conventional HDR-BT optimization, the RSBT optimization problem has the additional optimization variables of radiation exposure time at each angle of the shield. Due to the increased degrees of freedom in RSBT, RSBT optimization is more difficult than that for the conventional HDR-BT. In addition, there is a compelling need to quickly obtain optimal treatment plans in RSBT to enable clinical usage. To achieve this, researchers have used the dose-surface optimization (DSO) method\cite{liu2013rapid,yang2013rotating}, which minimizes the total dose errors over only voxels on the HR-CTV surface. Instead of dealing with only voxels on the HR-CTV surface, Liu \textit{et al.}\cite{liu2014asymmetric} defined the region of interest in tumor which includes the surface of HR-CTV as well as the inside voxels of tumor. Additionally, the authors used the total variation (TV) norm penalty in their optimization problem to make smooth changes in the emission times of adjacent beams in the treatment process to facilitate the efficient delivery of an RSBT plan. This optimization problem for RSBT is called asymmetric dose-volume optimization with smoothness control (ADOS).

In this paper, we consider the ADOS optimization problem. A fast computational method is proposed to solve the ADOS optimization problem for the optimal cancer treatment planning for RSBT. Liu \textit{et al.} used a commercial optimization solver called CPLEX\cite{liu2014asymmetric}. In order to efficiently solve the ADOS optimization problem, which is a large-scale RSBT optimization problem, we designed an optimization method based on the proximal graph solver (POGS)\cite{pogs}, which is a solver using the alternating direction method of multipliers (ADMM). For using POGS, we derived closed-form formulas for the proximal operators used in POGS. Further, we applied our method to the H-RSBT, which is a mechanically-feasible delivery technique for RSBT proposed by Dadkhah \textit{et al.}\cite{dadkhah2015multihelix}. In the numerical experiments, we considered cervical cancer, even though our method is also applicable to other types of cancer such as breast cancer and prostate cancer.

\section{MATERIALS AND METHODS}
\label{sec:material}
\subsection{Delivery method}
\label{subsec:hrsbt}

\begin{figure*}[t]
    \centering
    \includegraphics[scale=0.23]{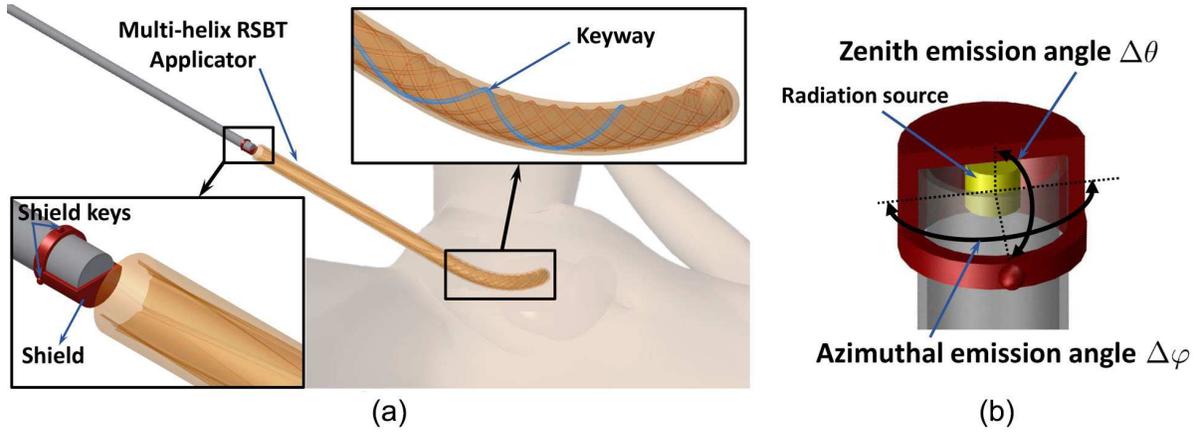}
    \caption{(a) Illustration of multihelix rotating shield brachytherapy (H-RSBT) system. (b) Partially shielded radiation source.}
    \label{fig:multihelix1}
\end{figure*}

In order to deliver the radiation dose to a target organ, we consider a mechanically-feasible delivery technique for RSBT, called the multi-helix RSBT (H-RSBT) \cite{dadkhah2015multihelix}. Fig. \ref{fig:multihelix1} shows the illustration of the H-RSBT method. The shield opening is represented by the \textit{azimuthal} and \textit{zenith} emission angles, denoted by $\Delta \varphi$ and $\Delta \theta$ respectively.

In H-RSBT, a Xoft Axxent radiation source, inside its cooling catheter, with a freely-rotating partial radiation shield is translated inside an applicator with six helical keyways carved out of the inner wall. The six keyways are evenly spaced on the applicator cross section, by $60\degree$, and each keyway has a helical pitch of one rotation per $33.3\; mm$ of translation. The partial shield has a protruding key that travels down a given keyway, and, due to the helical design of the keyways, the shield rotates about the radiation source as the source catheter is translated, and the emission angle of the shield is known for a given keyway and translational dwell position. As the H-RSBT applicator has 6 helices, with $33.3\; mm$ of translation along the applicator per helical rotation and $1.7\; mm$ spacing between dwell positions, it yields 17.5\degree of rotation for the shield per $5\; mm$ (standard dwell position spacing) of its translation along the applicator. The dose calculation resolution was $1 \times 1 \times 3\; mm^3$ for all cases. The transmission through the shield is 0.1\% and approximated to be 0\% for the dose calculation. For each patient considered, 45\degree azimuthal emission angle was used for treatment planning. The zenith emission angle of the modeled shields was an asymmetric 116\degree, which is consistent with previous work.\cite{dadkhah2015multihelix}

\subsection{Radiation source model and dose calculation}

For H-RSBT, the delivery is parameterized by keyway number and dwell position number along the keyway. To quantitatively describe the structure of high-dose-regions formed by a partially shielded source, we introduce the notation of {\em beamlet}. A beamlet, denoted by $D_i(j,b)$, is defined as the dose rate at point $\vec{r}_i$ with the shielded source positioned at the $j$-th dwell position while the shield is aligned with the $b$-th keyway.

To calculate the beamlet, we use the TG-43 dose calculation model of Rivard \textit{et al.}\cite{rivard2006calculated}. The radiation source is assumed to be partially shielded 50kVp Xoft Axxent$^{TM}$ (Sunnyvale, CA). To be consistent with previous work\cite{liu2013rapid,yang2013rotating,velpula2012computational}, we consider that the dose to the points blocked by the shield is 0, since the transmission rate from 50kVp Xoft Axxent$^{TM}$ can be controlled to be less than 0.1\% when using a $0.5\;mm$ tungsten shield.\cite{yang2013rotating,liu2013rapid} Then, we can quantify the radiation dose amount at the point $\vec{r}_i$, denoted by $d_i$, as a time-weighted sum of all beamlets as follows:
\begin{align}
\label{eq:beamlet}
          d_i = \sum_{j,b} D_i(j,b) t_{j,b},
\end{align}
where $t_{j,b}$ is the duration time of the beamlet $D_i(j,b)$.

In the next subsection, we introduce the RSBT optimization problem having asymmetric penalty parameters for HR-CTV and organs at risk (OARs), with TV regulation term for smoothness in the beamlet emission times.

\subsection{Optimization problem for cancer treatment planning in RSBT}
\label{subsec:opt_prob}
\subsubsection{Problem formulation}

Let $t \in \mathbb{R}^{mn \times 1}$ be the beamlet emission time vector for all keyways and all dwell positions, where $m$ and $n$ are the number of keyways, i.e, $m=6$, and the number of dwell positions along a keyway respectively. We can obtain $t$ by vectorizing $t_{j,b}$ in Eqn. (\ref{eq:beamlet}); namely, the vector $t \in \mathbb{R}^{mn \times 1}$ is a concatenated vector, which is expressed as $t = [t[1]^T, t[2]^T, ..., t[m]^T]^T$, where $t[j] \in \mathbb{R}^{n \times 1}$ is the dwell time vector for all the beamlets along a keyway, and the super-script $T$ represents the transpose. Let us define a concatenated dose rate matrix $D = [D[1], D[2], ..., D[m]] \in \mathbb{R}^{l \times mn}$, where $D[j] \in \mathbb{R}^{l \times n}$, $j=1,...,m$, is the dose rate matrix for the $j$-th selected keyway, and $l$ is the number of voxels that we are interested in. We denote the whole index set for voxels of interest (VOIs) as $\mathcal{I}_{VOIs}$ and the index set for HR-CTV, bladder, rectum, sigmoid, and normal tissue around the HR-CTV as $\mathcal{I}_{tumor}$, $\mathcal{I}_{bladder}$, $\mathcal{I}_{rectum}$, $\mathcal{I}_{sigmoid}$, and $\mathcal{I}_{normal}$ respectively. The dose rate matrix $D$ has information about dose rate delivered to each tissue point in $\mathcal{I}_{VOIs}$ from each beamlet.

We consider the following RSBT optimization problem having a quadratic objective function with total variation (TV) regulation term for smooth beamlet emission times:
\begin{align}
\label{eq:TV_min}
          & \underset{t \in \mathbb{R}^{mn},d\in \mathbb{R}^l}{\text{minimize}}  \quad \sum_{i \in \mathcal{I}_{VOIs}} h(d_i) + \sum_{j=1}^{m}\beta||Lt[j]||_1 \nonumber \\
          & \text{subject to}          \quad Dt = d,  \nonumber \\
          &                            \quad\quad\quad\quad\quad\;\; t \geq 0,
\end{align}
where $D \in \mathbb{R}^{l \times mn}$ is a dose rate matrix, $t \geq 0$ is the element-wise non-negative emission time, $d_i$, which is the $i$-th element of $d$, is the dose amount at the $i$-th voxel as introduced in Eqn.\,(\ref{eq:beamlet}). In  Eqn.\,(\ref{eq:TV_min}),
    \[h(d_i) \triangleq (\lambda_i^{+}H(d_i - \hat{d_i}) + \lambda_i^{-} H(\hat{d_i} - d_i))(d_i - \hat{d_i})^2,\]
where $H(x)$ is the unit step function, which is $H(a)=1$ if $a >0$, and $H(a)=0$ if $a \leq 0$, and $\lambda_i^{+}$ and $\lambda_i^{-}$ represent overdose and underdose penalty parameters for the $i$-th voxel respectively. $L \in \mathbb{R}^{n \times n}$ is the matrix which calculates TV norm of a vector; namely, $L$ is defined as follows:
\begin{align*}
    L \triangleq \begin{bmatrix}
                 1      & -1     & 0      & 0      & ... & 0 & 0\\
                 0      & 1      &-1      & 0      & ... & 0 & 0\\
                 0      & 0      & 1      &-1      & ... & 0 & 0\\
                 \vdots & \vdots & \vdots & \vdots & \vdots & \vdots &\vdots\\
                 0      & 0      & 0      & 0      & ... & 1 & -1\\
                 0      & 0      & 0      & 0      & ... & 0 & 0
                 \end{bmatrix}.
\end{align*}
$\hat{d_i}$ is a prescribed dose amount for the $i$-th voxel. $\hat{d_i}$ can have a different value for each VOI. For example, $\hat{d_i} = \hat{d}_{tumor}$ if $i \in \mathcal{I}_{tumor}$, and $\hat{d_i} = \hat{d}_{badder}$ if $i \in \mathcal{I}_{bladder}$.
We denote the prescribed dose amount for HR-CTV, bladder, rectum, sigmoid, and normal tissue around the HR-CTV as $\hat{d}_{tumor}$, $\hat{d}_{bladder}$, $\hat{d}_{rectum}$, $\hat{d}_{sigmoid}$, and $\hat{d}_{normal}$. For $\lambda_i^{+}$ and $\lambda_i^{-}$, we use different non-negative overdose and underdose parameter values including 0.

The TV norm alleviates the positioning uncertainty in the treatment process. If we have two dramatically different emission times $t_{j,b}$ and $t_{j+1,b}$ between two adjacent beamlets along the same keyway, a small error in the dwell positions may cause an unacceptable treatment result. By applying the smoothness term between two adjacent beamlets along the same keyway in H-RSBT, we can reduce the treatment error caused by the positioning uncertainty in the treatment process.

Since we use different penalty parameter values for the overdose and underdose of a voxel, we call Eqn. (\ref{eq:TV_min}) as RSBT optimization problem having asymmetric penalty parameters or simply ADOS.

\subsubsection{POGS implementation}

In order to simplify the sum of the TV norms in Eqn. (\ref{eq:TV_min}), let us introduce a matrix $\bar{L} \triangleq I_{m \times m} \otimes L$, where $\otimes$ is the Kronecker product, and $I_{m \times m}$ is an $m \times m$ identity matrix. By assigning $\bar{L}t = y$ and introducing the indicator function $I(\cdot)$, we restate Eqn.\,(\ref{eq:TV_min}) as follows:
\begin{align}
\label{TV_min_ADMM2}
          &\underset{t,y \in \mathbb{R}^{mn}, d \in \mathbb{R}^{l}}{\text{minimize}}   \quad \sum_{i \in \mathcal{I}_{VOIs} } h(d_i) + \beta||y||_1 + I(x \geq 0) \nonumber \\
          &\;\;\;\text{subject to}     \quad Dt = d, \nonumber \\
          &                           \;\;\;\quad\quad\quad\quad\quad\;\; \bar{L}t = y,
\end{align}
where $D \in \mathbb{R}^{l \times mn}$, $\bar{L} \in \mathbb{R}^{mn \times mn}$, and $I(t \geq 0)$ is the element-wise indicator function; namely, $I(t_i \geq 0)=0$ if $t_i \geq 0$, and $I(t_i \geq 0) = \infty$ if $t_i < 0$.

By letting
\begin{align*}
    A = \begin{pmatrix}D\\\bar{L} \end{pmatrix},\;\; z = \begin{pmatrix} d\\y \end{pmatrix},
\end{align*}
we can further simplify Eqn.\,(\ref{TV_min_ADMM2}) into
\begin{align}
\label{TV_min_ADMM3}
          &\underset{t \in \mathbb{R}^{mn}, z \in \mathbb{R}^{l+mn} }{\text{minimize}}   \quad \sum_{i=1}^{l} h(z_i)  + \beta||z_{[l+1:l+mn]}||_1 + I(t \geq 0) \nonumber \\
          &\quad\text{subject to}           \quad At = z,
\end{align}
where $z_{[a:b]}$ is the partial vector of $z$ by taking vector $z$ from the $a$-th element to the $b$-th element. We define the following functions:
\begin{align}
    & g(t) = I(t \geq 0),\\
    & f(z) =  \sum_{i=1}^{l} h(z_i) + \beta||z_{[l+1:l+mn]}||_1.
\end{align}
Then we turn Eqn.\,(\ref{TV_min_ADMM3}) into a graph-form convex optimization problem\cite{pogs}, where the constraint is $z=At$, and $A=[D^T \bar{L}^T]^T \in \mathbb{R}^{(l+mn) \times mn}$. We have derived the detailed updating rules for each optimization variable in the POGS solver for Eqn.\,(\ref{TV_min_ADMM3}). We introduce our derived results in detail for the proximal operators used in the POGS solver, updating steps, and stopping criteria as follows.

The POGS updates primal variables, conducts the projection onto the space $z=At$, and then, updates dual variables iteratively until the stopping criteria are satisfied or the maximum number of iterations, denoted by $MaxItr$, is reached. The primal variable and dual variable are updating variables to be used for optimality condition in the algorithm. For the primal variable, dual variable, and projection result, we used $(\hat{t}, \hat{z})$, $(t,z)$, and $(\tilde{t},\tilde{z})$ respectively. We introduce each updating steps in detail for our optimization problem in Eqn. (\ref{TV_min_ADMM3}). We use the super-script $k$ to represent the $k$-th iteration.

\textbf{Updating primal variables $\hat{t}^{k+1}$ and $\hat{z}^{k+1}$:} In updating the primal variables, we use the following proximal operators with a penalty parameter $\rho$:
\begin{align*}
      \hat{t}^{k+1} & = Prox_g(t^k - \tilde{t}^k)  \\
                & = \underset{t}{argmin}\;\; I(t \geq 0) + \frac{\rho}{2}||t - (t^k - \tilde{t}^k)||^2,\\
      \hat{z}^{k+1} & = Prox_f(z^k - \tilde{z}^k)  \\
                & = \underset{z}{argmin}\;\; \sum_{i=1}^{l} h(z_i) + \beta||z_{[l+1:l+mn]}||_1 \\
                & \quad\quad\quad\quad\quad + \frac{\rho}{2}||z - (z^k - \tilde{z}^k)||^2.
\end{align*}
The proximal operator is used to make a compromise between the solution at the $k$-th iteration and the function value with the solution at the $k+1$ iteration. We are able to explicitly derive closed-form formulas for the proximal operators. For $\hat{t}^{k+1} \in \mathbb{R}^{mn}$, we have
\begin{align}
\label{eq:primal_x}
    \hat{t}^{k+1} = max(t^k - \tilde{t}^k,\textbf{0}),
\end{align}
where $max(a,b)$ provides the maximum value between $a$ and $b$ element-wise. For $\hat{z}^{k+1}_i$, $1 \leq i \leq l$, we also derive
\begin{align}
\label{eq:primal_z1}
    \hat{z}^{k+1}_i  = \begin{cases}
                    & z^k_i - \tilde{z}^k_i - \frac{\beta}{\rho},\;\; \text{if}\; z^k_i - \tilde{z}^k_i > \frac{\beta}{\rho} \\
                    & z^k_i - \tilde{z}^k_i + \frac{\beta}{\rho},\;\; \text{if}\; z^k_i - \tilde{z}^k_i < - \frac{\beta}{\rho}\\
                    & 0 ,\;\; \text{otherwise}
      \end{cases}.
\end{align}
For $\hat{z}^{k+1}_i$, $l+1 \leq i \leq l+mn$, we obtain
\begin{align}
\label{eq:primal_z2}
 \hat{z}^{k+1}_i = \begin{cases}
                   & \frac{2 \lambda^{+}_i \hat{d}_i + \rho (z^{k}_i - \tilde{z}^k_i)}{2 \lambda^{+}_i + \rho}, \;\;\text{if}\;\;z^{k}_i - \tilde{z}_i^{k} \geq \hat{d}_i, \\
                   & \frac{2 \lambda^{-}_i \hat{d}_i + \rho (z^{k}_i - \tilde{z}^k_i)}{2 \lambda^{-}_i + \rho}, \;\;\text{if}\;\;z^{k}_i - \tilde{z}_i^{k} < \hat{d}_i.
                \end{cases}.
\end{align}
POGS uses the adaptive value for $\rho$ as default to further increase the convergence speed.

\textbf{Projection onto $z=At$ from $(\hat{t}^{k+1}+\tilde{t}^{k},\hat{z}^{k+1}+\tilde{z}^k)$:} The projection operation is mapping the primal variables to the closest feasible solution. The projected variables onto $z=At$ from $(\hat{t}^{k+1}+\tilde{t}^{k},\hat{z}^{k+1}+\tilde{z}^k)$, denoted as $t^{k+1}$ and $z^{k+1}$, are obtained by solving the following optimization:
\begin{align*}
          &\underset{t,z}{\text{minimize}}   \quad \frac{1}{2}|| t - (\hat{t}^{k+1}+\tilde{t}^{k})||^2_2 +  \frac{1}{2}|| z - (\hat{z}^{k+1}+\tilde{z}^k)||^2_2 \nonumber \\
          &\text{subject to} \quad    At=z
\end{align*}
By solving this optimization and using Lagrange conditions\cite{boyd2004convex}, we have the following formulation:
\begin{align*}
    \begin{pmatrix}t^{k+1}\\z^{k+1}\end{pmatrix} = \begin{pmatrix}I & A^T \\A & -I\end{pmatrix}^{-1}  \begin{pmatrix}\hat{t}^{k+1} + \tilde{t}^{k} + A^T (\hat{z}^{k+1} + \tilde{z}^{t}) \\ 0 \end{pmatrix}.
\end{align*}

\textbf{Updating dual variables $\tilde{t}^{k+1}$ and $\tilde{z}^{k+1}$:} We obtain the dual variable at iteration $(k+1)$ by updating the dual variable at the $k$-th iteration as follows:
\begin{align*}
    & \tilde{t}^{k+1} = \tilde{t}^{k} + \hat{t}^{k+1} - t^{k+1}, \\
    & \tilde{z}^{k+1} = \tilde{z}^{k} + \hat{z}^{k+1} - z^{k+1}.
\end{align*}

We summarize the updating steps in Algorithm \ref{alg:POGS}.

\textbf{Stopping criteria:} For the stopping criteria, we define the primal and dual residuals as follows:
\begin{align}
    & ||A\hat{t}^{k+1} - \hat{z}^{k+1}||_2 \leq \epsilon^{pri}, \nonumber \\
    & ||A^T \hat{v}^{k+1} + \hat{\mu}^{k+1}||_2 \leq \epsilon^{dual},
\end{align}
where $\hat{v}^{k+1}=-\rho(\hat{z}^{k+1}-z^{k} + \tilde{z}^{k})$, $\hat{\mu}^{k+1}=-\rho(\hat{t}^{k+1}- t^{k} + \tilde{t}^{k})$. Here, $\epsilon^{pri}$ and $\epsilon^{dual}$ are positive tolerances for primal and dual residuals respectively:
\begin{align}
    & \epsilon^{pri} = \epsilon^{abs} + \epsilon^{rel}||\hat{z}^{k+1}||_2, \nonumber \\
    & \epsilon^{dual}=\epsilon^{abs} + \epsilon^{rel}||\hat{\mu}^{k+1}||_2,
\end{align}
where we used $(\epsilon^{abs},\epsilon^{rel}) = (10^{-4}, 10^{-2})$ in the numerical experiments.

\begin{algorithm}[t]
  \caption{Fast treatment planning for RSBT in POGS implementation}
  \label{alg:POGS}
\SetAlgoNoLine
{\footnotesize
   \KwIn{ $A \in \mathbb{R}^{(l+mn)\times mn}$, \textit{MaxItr}, $\lambda^{+},\lambda^{-},\hat{d} \in \mathbb{R}^{l}$, $\beta$ }
   \KwOut{ $t$ }
   \textbf{Initialize:} $k \leftarrow 0$, $t^{k} \leftarrow 0$, $z^{k} \leftarrow 0$, $\tilde{t} \leftarrow 0$,$\tilde{z} \leftarrow 0$  \par
   \For { $k=1$  \KwTo MaxItr }
   {
       Updating primal variables $\hat{t}^{k+1}$, $\hat{z}^{k+1}$: \par
            \quad $\hat{t}^{k+1} \leftarrow Prox_g(t^k - \tilde{t}^k) $ \Comment{See (\ref{eq:primal_x})}\par
            \quad $\hat{z}^{k+1} \leftarrow Prox_f(z^k - \tilde{z}^k) $ \Comment{See (\ref{eq:primal_z1}) and (\ref{eq:primal_z2})}\par
       Projection onto $z=At$: \par
            \quad
             $\begin{pmatrix}t^{k+1}\\z^{k+1}\end{pmatrix} \leftarrow \begin{pmatrix}I & A^T \\A & -I\end{pmatrix}^{-1}  \begin{pmatrix}\hat{t}^{k+1} + \tilde{t}^{k} + A^T (\hat{z}^{k+1} + \tilde{z}^{t}) \\ 0 \end{pmatrix}$ \par
      Updating dual variables $\tilde{t}^{k+1}$, $\tilde{z}^{k+1}$: \par
            \quad $\tilde{t}^{k+1} \leftarrow \tilde{t}^{k} + \hat{t}^{k+1} - t^{k+1}$ \par
            \quad $\tilde{z}^{k+1} \leftarrow \tilde{z}^{k} + \hat{z}^{k+1} - z^{k+1}$ \par
       \If {Stopping criteria are met}
      {
      \quad break
      }
   }
}%
\end{algorithm}

\subsection{Treatment planning}
\label{subsubsec:patient}
Five patients with cervical cancer were considered, whose HR-CTV volumes range from $42.2$ to $98.8\;cm^3$. Table \ref{tbl:patient} shows the volume and maximum dimension of HR-CTV for five patients. All the HR-CTVs and OARs were manually contoured by physicians on T2 weighted $1 \times 1 \times 3$ $mm^3$ resolution MR images taken with a Siemens MAGNETOM 3T scanner (Siemens, Germany) at the beginning of the first fraction of brachytherapy. A titanium Fletcher-Suit-Delclos style tandem and ovoids (Varian Medical Systems, Palo Alto, CA) were used as the brachytherapy applicator. We used the same datasets as the previous research conducted by Liu \textit{et al.}\cite{liu2014asymmetric} and Dadkhah \textit{et al.}\cite{dadkhah2015multihelix}.

\begin{table}[h!]
\centering
\caption{\label{tbl:patient} HR-CTV volumes and dimensions for all patients}
\begin{threeparttable}
\begin{ruledtabular}
\small{
\begin{tabular}{c p{2cm}p{3cm}}
Patient Num. & HR-CTV volume ($cm^3$) & HR-CTV maximum dimension ($cm$) \\ \hline
       Case 1 &   \quad\quad 42.2   & \quad\quad\quad 6.3      \\
       Case 2 &   \quad\quad 45.8   & \quad\quad\quad 7.4      \\
       Case 3 &   \quad\quad 78.0   & \quad\quad\quad 8.6      \\
       Case 4 &   \quad\quad 98.8   & \quad\quad\quad 9.6      \\
       Case 5 &   \quad\quad 75.0   & \quad\quad\quad 7.5      \\
       Avg. &     \quad\quad 68.0   & \quad\quad\quad 7.9      \\
       SD\tnote{a} & \quad\quad 23.8 & \quad\quad\quad 1.8
\end{tabular}
}
\end{ruledtabular}
\begin{tablenotes}
\setlength\itemsep{-0.5pt plus 1pt minus 1pt}
    \item [a] {\scriptsize Standard Deviation}
\end{tablenotes}
\end{threeparttable}
\end{table}

\begin{table*}[t]
\centering
\caption{\label{tbl:para_setting} Parameter settings}
\begin{threeparttable}
\setlength{\tabcolsep}{4.5pt}
\footnotesize{
\begin{tabular}{p{2cm} c c c c c c c c c c c}
\hline\hline
  \multirow{2}{*}{Method} & \multirow{2}{*}{ $\hat{d}_{tumor}$ } & \multirow{2}{*}{ $\hat{d}_{bladder}$ } & \multirow{2}{*}{ $\hat{d}_{rectum}$ }  & \multirow{2}{*}{ $\hat{d}_{sigmoid}$ } & \multirow{2}{*}{ $\hat{d}_{normal}$\tnote{a} }& Tumor & Bladder & Rectum & Sigmoid  & Normal\tnote{b}  & \multirow{2}{*}{$\beta$} \\ \cline{7-11}
                        &     &     &    &    &       & $\lambda_i^{+}$/ $\lambda_i^{-}$ & $\lambda_i^{+}$/ $\lambda_i^{-}$    & $\lambda_i^{+}$/ $\lambda_i^{-}$   & $\lambda_i^{+}$/ $\lambda_i^{-}$   & $\lambda_i^{+}$/ $\lambda_i^{-}$    &        \\ \hline
 CPLEX                  & 40  & 25  & 20 & 20 &  40   & 0/ 2  & 2/ 0     & 2/ 0    & 2/ 0    & 2/ 0     & 100    \\
 POGS                   & 40  & 25  & 20 & 20 &  40   & 0/ 2  & 2/ 0     & 2/ 0    & 2/ 0    & 2/ 0     & 100    \\ \hline\hline
\end{tabular}
}
\begin{tablenotes}
\setlength\itemsep{-0.5pt plus 1pt minus 1pt}
    \item [a] {\scriptsize Prescribed dose amount for tumor boundary}
    \item [b] {\scriptsize Penalty parameter for tumor boundary }
\end{tablenotes}
\end{threeparttable}
\end{table*}

\begin{table*}[t]
\centering
\caption{\label{tbl:comparison_hrsbt} Comparison between POGS and CPLEX for 45 \degree azimuthal angle}
\begin{threeparttable}
\setlength{\tabcolsep}{10pt}
\small{
\begin{tabular}{p{1cm} p{2.5cm} c c c c c c c }
\hline\hline
  \multirow{2}{*}{Case}  & \multirow{2}{*}{ Method } & HR-CTV & HR-CTV  & Bladder & Rectum & Sigmoid  & Execution  \\
&     & $D_{90}$ (Gy)  & $D_{100}$ (Gy)     & $D_{2cc}$ (Gy) & $D_{2cc}$ (Gy) & $D_{2cc}$ (Gy) & time (sec.) \\ \hline
\multirow{2}{*}{Case 1 } & CPLEX  & 110.8 & 54.0 & 89.9 & 62.4 & 75.0 & 32.1    \\
                         & POGS   & 111.4 & 54.0 & 90.0 & 64.7 & 74.7 & 2.1    \\ \hline
\multirow{2}{*}{Case 2 } & CPLEX  & 111.5 & 44.3 & 90.0 & 72.2 & 54.4 & 37.0    \\
                         & POGS   & 111.5 & 44.3 & 90.0 & 71.7 & 54.8 & 2.1    \\ \hline
\multirow{2}{*}{Case 3 } & CPLEX  & 96.0  & 44.3 & 85.9 & 57.3 & 75.0 & 65.4   \\
                         & POGS   & 95.0  & 44.3 & 85.2 & 55.1 & 75.0 & 3.9     \\ \hline
\multirow{2}{*}{Case 4 } & CPLEX  & 107.0 & 55.3 & 90.0 & 69.9 & 54.0 & 39.4    \\
                         & POGS   & 106.9 & 55.4 & 90.0 & 69.8 & 54.0 & 2.3     \\ \hline
\multirow{2}{*}{Case 5 } & CPLEX  & 112.7 & 44.3 & 90.0 & 68.1 & 59.2 & 65.4    \\
                         & POGS   & 112.7 & 44.3 & 90.0 & 68.1 & 59.2 & 3.2     \\ \hline
\multirow{2}{*}{Average} & CPLEX  & 107.6 & 48.4 & 89.2 & 66.0 & 63.5 & 47.9  \\
                         & POGS   & 107.5 & 48.5 & 89.0 & 65.9 & 63.5 & 2.7   \\
                        \hline\hline
\end{tabular}
}
\end{threeparttable}
\end{table*}

\begin{figure*}[t]
    \centering
    \includegraphics[scale=1.5]{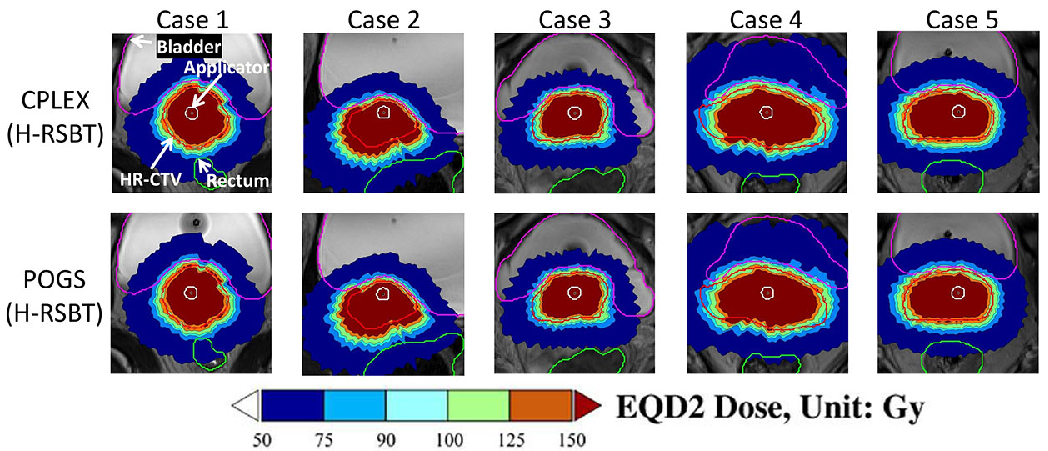}
    \caption{EQD2 dose distributions on MR images for five patient cases obtained from CPLEX and POGS with H-RSBT using 45\degree azimuthal angle.}
    \label{fig:EQD2_hrsbt}
\end{figure*}

All the patients had external beam radiation treatment in 25 fractions at 1.8 Gy/fraction. We assume that the external beam radiotherapy dose was uniformly delivered to the HR-CTV and OARs for all the patient cases. The dose in each voxel was converted to equivalent dose in 2 Gy per fraction of external radiation therapy (EQD2) using the linear quadratic model,\cite{joiner2016basic} where the linear-quadratic parameter, $\nicefrac{\alpha}{\beta}$, is set to 3 Gy for OARs and 10 Gy for HR-CTV.

For VOIs, we define the voxels located at a distance between $3\; mm$ and $20\; mm$ to the radiation source path or those within $10\; mm$ inside and outside of the HR-CTV boundary surface.\cite{liu2014asymmetric} We deal with the HR-CTV, HR-CTV boundary, bladder, sigmoid, and rectum inside of VOIs in our optimization problem. The optimization parameter settings are shown in Table \ref{tbl:para_setting}.

For all the brachytherapy treatment plans, we escalated the EQD2 of the HR-CTV without exceeding the $D_{2cc}$ tolerance of the bladder, rectum, and sigmoid colon. We used 90 Gy for bladder tolerance, and 75 Gy for rectum and sigmoid colon tolerances according to Groupe Europ\'een de Curieth\'erapie, European Society for Therapeutic Radiology and Oncology (GEC ESTRO).\cite{potter2006recommendations,haie2005recommendations}


\subsection{Treatment plan evaluation}
Optimized treatment plans were generated for all patients using the POGS method and the previously considered CPLEX method.\cite{liu2014asymmetric} The same objective function, with the same input parameters and beamlets, was minimized for each patient with both methods. A total variation term was included in the objective function as a regularization term, resulting in smoothly-varying emission times along each keyway. The regularization promotes robustness of the resulting overall dose distribution with respect to small errors (expected $\leq 1\; mm$) in source positioning. The rectum, bladder, sigmoid colon, HR-CTV, and HR-CTV boundary were the structures considered.

We compared our method with the previous research conducted by Liu \textit{et al.}\cite{liu2014asymmetric} using CPLEX\cite{cplex} for H-RSBT. We evaluated the quality of the delivery plans as well as the execution time to solve Eqn.~\,(\ref{eq:TV_min}) with POGS\cite{pogs}. Since Liu \textit{et al.} compared their method based on CPLEX with other existing RSBT dose optimization methods ranging from DSO to IPSA in their previous research\cite{liu2014asymmetric}, we only compared POGS and CPLEX in this paper.

The comparison metrics for the quality of the delivery plans are the HR-CTV $D_{90}$, HR-CTV $D_{100}$, OARs $D_{2cc}$, DVH, and dose distributions. Since the goal of this research is achieving a fast solution to the RSBT dose optimization problem without compromising the plan quality, we compared the execution times to solve Eqn.~\,(\ref{eq:TV_min}) for all five patient cases. We conducted our numerical experiments on HP Z220 CMT with an Intel Core i7-3770 dual core CPU @3.4GHz clock speed and 16GB DDR3 RAM, using Matlab (R2013b) on the Windows 7 operating system.

\begin{table}[h!]
\centering
\caption{\label{tbl:D_size} Dimension of $D \in \mathbb{R}^{l \times mn}$ in Eqn. (\ref{eq:TV_min})}
\begin{threeparttable}
\begin{ruledtabular}
\small{
\setlength{\tabcolsep}{5pt}
\begin{tabular}{c p{4cm}}
Patient Num.  &   $l \times mn$   \\ \hline
       Case 1 &   $54693 \times 144$  \\
       Case 2 &   $51109 \times 126$  \\
       Case 3 &   $79065 \times 222$  \\
       Case 4 &   $50680 \times 144$  \\
       Case 5 &   $59220 \times 228$  \\
       Avg.   &   $58953 \times 173$  \\
\end{tabular}
}
\end{ruledtabular}
\end{threeparttable}
\end{table}


\section{RESULTS}

\begin{figure*}[t]
    \centering
    \includegraphics[scale=1.8]{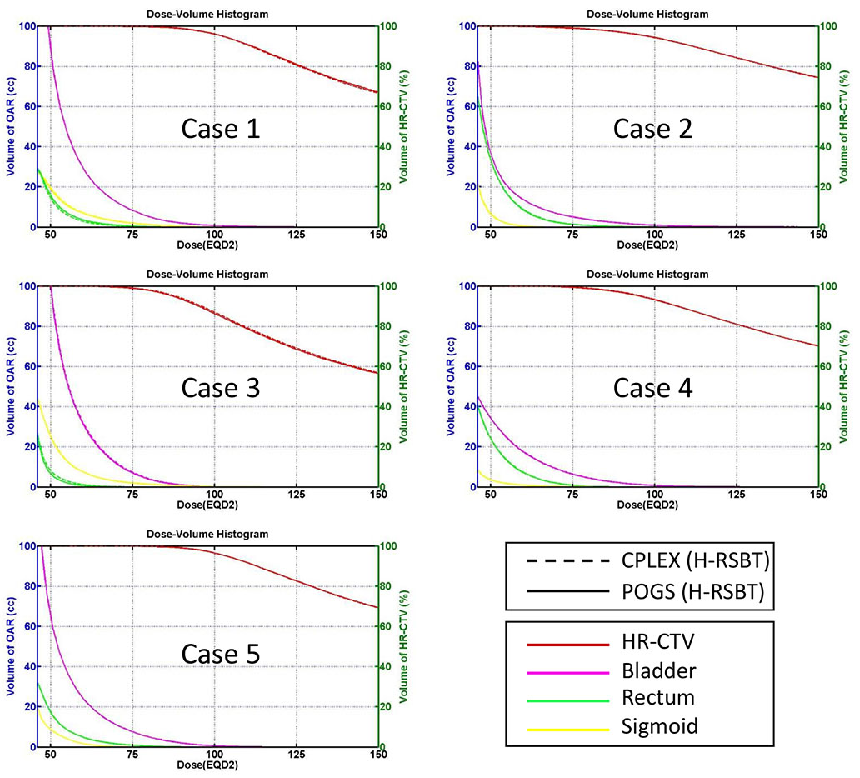}
    \caption{Dose volumn histograms (DVH) of all treatment planning for five patient cases in H-RSBT with 45\degree azimuthal angle.}
    \label{fig:DVH_hrsbt}
\end{figure*}


Results for five cervical cancer patient cases are shown in Table \ref{tbl:comparison_hrsbt}. Table~\ref{tbl:D_size} shows the dimension of the dose matrix $D$ in Eqn.~\,(\ref{eq:TV_min}). Fig. \ref{fig:DVH_hrsbt} shows the corresponding DVH in H-RSBT. With the same parameter settings as in Table \ref{tbl:para_setting}, POGS can achieve an RSBT plan with almost the same $D_{90}$ value (less than 1\% difference) as that achieved by CPLEX in each of the five patient cases. $D_{2cc}$ values for OARs obtained by POGS are also almost the same (less than 1\% difference) as those from CPLEX. For the execution time, we achieved around 18 times faster speed to solve the ADOS problem for H-RSBT than the CPLEX based method on average. Over all patients, total optimization times were 32.1-65.4 seconds for CPLEX and 2.1-3.9 seconds for POGS.

Fig. \ref{fig:EQD2_hrsbt} shows that the EQD2 figures were similar for each case between CPLEX and POGS.

\section{DISCUSSION}
\label{sec:discussion}

Various treatment planning methods in radiation therapy have been studied. One of the well known methods, which is called IPSA, was introduced to directly optimize DVH with given constraints in heuristic way. Due to its heuristic nature, a global solution is not guaranteed. Instead of directly optimizing DVH, we consider the voxel-wise optimization problem for the RSBT treatment planning, called ADOS optimization problem, which can be expressed in a convex optimization problem. We can take advantage of the convexity to obtain a global solution.

In the ADOS optimization problem, we reduced the size of the ADOS optimization problem by defining VOIs. Instead of defining VOIs, the whole voxels can be considered in the optimization problem under the expectation of better treatment quality with heavy computation. Since parallel computing and GPU-based high performance computing can play an important role in solving extremely large-scale optimization problems, there is a rising question about the usability of POGS in parallel computing environment or GPU-based implantation. The implementation of POGS in such environment is another research work.

In addition, we used a partial shield with 45\degree azimuthal angle in H-RSBT for our numerical experiments. However, finding the optimal shield angle in H-RSBT is still an open problem. In order to determine the size of angle and the radiation exposure time at each angle of shield, considering both variables in the optimization problem is also a possible optimization problem.

Finally, POGS (and ADMM) was used in previous research on intensity-modulated radiation therapy (IMRT)\cite{pogs,Zarepisheh2017computation}, fluence map optimization\cite{Gao2016Robust}, and external beam radiotherapy (EBRT) optimization\cite{Liu2017Use}. Right at the time of submitting our journal manuscript, we learned of the recently-appearing (published on April 12th, 2017) work \cite{Liu2017Use} which applied POGS algorithms to EBRT dose optimization.  The work \cite{Liu2017Use} focused on EBRT,  while our paper is the first work to use POGS in brachytherapy, including the mechanically-feasible delivery system called H-RSBT. In our paper, we use TV norm to promote smoothly-varying emission times along each keyway, such that  the treatment plan is robust to positioning errors of dwell positions. By comparison, the TV norm is instead applied to promote the smoothness of the resulting fluence map in the research\cite{Liu2017Use} and simplify the delivery. Our proposed method is applicable to conventional HDR-BT as well as dynamic modulated brachytherapy\cite{webster2011dynamic} with simple modifications, since they share similar mechanisms as H-RSBT.

\section{CONCLUSIONS}
\label{sec:conclusion}
POGS substantially reduced treatment plan optimization time around 18 times for RSBT with similar HR-CTV $D_{90}$, OAR $D_{2cc}$ values, and EQD2 figure comparing to CPLEX, which is significant progress toward clinical translation of RSBT. Over all cervical cancer patients, total optimization times were 32.1-65.4 seconds for CPLEX and 2.1-3.9 seconds for POGS. POGS is also applicable to conventional high-dose-rate brachytherapy.


\section{ACKNOWLEDGEMENT}
The authors would like to thank the generous support from NIH 1R01EB020665 and NSF DMS-1418737. The authors would also like to thank Jianfeng Cai,  Qihang Lin, and Haiye Huo for discussions and advice on paper writing.

%

\end{document}